\pdfoutput=1
\documentclass[acmsmall,screen,nonacm]{acmart}
\usepackage{booktabs}
\usepackage{pifont}
\newcommand{\cmark}{\ding{51}}
\newcommand{\xmark}{\ding{55}}
\usepackage{calc}
\usepackage{amsthm}
\newtheorem{definition}{Definition}

\usepackage{multirow}
\usepackage{dsfont}
\usepackage{enumitem}
\usepackage{algorithm}
\usepackage{algpseudocode}
\usepackage{listings}
\usepackage{xcolor}
\usepackage[table]{xcolor} 
\usepackage{tcolorbox}
\tcbuselibrary{listings,skins}
\usepackage{tikz}
\usetikzlibrary{arrows.meta,positioning,fit,backgrounds,shadows,decorations.pathreplacing,decorations.pathmorphing}
\usepackage{graphicx}
\usepackage{subcaption}
\usepackage{varwidth}
\usepackage{float}
\usepackage{placeins}
\usepackage{needspace}
\usepackage[frozencache]{minted}
\setminted{
  fontsize=\scriptsize,
  breaklines,
  tabsize=2,
  frame=none,
}
\usepackage{afterpage}

\newcommand{\program}{P}
\newcommand{\approachname}{Typed Template Fuzzing}
\newcommand{\approachnameShorthand}{TTF}
\newcommand{\toolname}{\textsf{Typos}}
\newcommand{\lengthOfBugHunting}{6}

\newcommand{\bughuntingMonths}{of 6 months}
\newcommand{\reportedBugs}{135}
\newcommand{\duplicateBugs}{7}
\newcommand{\unknownBugs}{128}
\newcommand{\confirmedBugs}{107}
\newcommand{\fixedBugs}{74}

\newcommand{\numInputLanguages}{21}
\newcommand{\numTargets}{27}
\newcommand{\allTargets}{GCC, G++, Clang, Clang++, rustc, Go, gccgo, javac, scalac, JerryScript, cvc5, Z3, slang, solc, Roslyn (csc), erlc, kotlinc, Lua, Perl, PHP, CPython, DuckDB, SQLite, scriptc, tsc, PyTorch and Qiskit}
\newcommand{\validityRate}{96\%}

\newcommand{\TotalCost}{215.45}
\newcommand{\TotalSeeds}{54,720}

\newcommand{\rqOne}{How effective is \toolname{} at finding bugs?}
\newcommand{\rqTwo}{Why is \toolname{} effective?}
\newcommand{\rqThree}{How does \toolname{} compare to other Fuzzers?}

\newcommand{\etc}{\emph{etc.}}

\newcommand{\ttemplate}[1]{\mathds{#1}}                       
\newcommand{\sort}[1]{\mathit{sort}(#1)}                      
\newcommand{\body}[1]{\mathit{body}(#1)}                      
\newcommand{\freevars}[1]{\mathit{free}(#1)}                      
\newcommand{\live}[1]{\mathit{live}(#1)}                      
\newcommand{\content}[1]{\mathit{text}(#1)}
\newcommand{\type}[1]{\mathit{type}(#1)}                      
\newcommand{\operands}{\mathit{operands}}                                 

\newcommand{\subst}[3]{#1[#2 / #3]}                         

\definecolor{tagleaf}{RGB}{110,160,215}
\definecolor{tagleafbg}{RGB}{230,240,251}
\definecolor{substtext}{RGB}{34,134,58}
\definecolor{substbg}{RGB}{230,255,237}
\newcommand{\markA}[1]{{\tiny\tcbox[on line, colframe=tagleaf, colback=tagleafbg, boxrule=0.4pt, arc=2pt, boxsep=0pt, left=1pt, right=1pt, top=0.5pt, bottom=0.5pt]{#1}}}
\newcommand{\markB}[1]{{\tiny\tcbox[on line, colframe=tagleaf, colback=tagleafbg, boxrule=0.4pt, arc=2pt, boxsep=0pt, left=1.5pt, right=1.5pt, top=1pt, bottom=1pt]{#1}}}
\newcommand{\markS}[1]{{\setlength{\fboxsep}{1pt}\colorbox{substbg}{\color{substtext}#1}}}
\definecolor{origseg}{RGB}{255,228,170}
\newcommand{\markR}[1]{{\setlength{\fboxsep}{0.3pt}\colorbox{origseg}{#1}}}
\newcommand{\markE}[1]{\tikz[baseline=(e.base)]{\node[inner sep=0pt, outer sep=0pt, text=red!80!black] (e) {#1};\draw[red!80!black, line width=.5pt, decorate, decoration={snake, amplitude=.6pt, segment length=2.4pt}] ([yshift=-1.5pt]e.base west) -- ([yshift=-1.5pt]e.base east);}}
\newcommand{\markAt}[1]{{\small\ttfamily\tcbox[on line, colframe=tagleaf, colback=tagleafbg, boxrule=0.4pt, arc=2pt, boxsep=0pt, left=1pt, right=1pt, top=0.5pt, bottom=0.5pt]{#1}}}
\newcommand{\markBt}[1]{{\small\ttfamily\tcbox[on line, colframe=tagleaf, colback=tagleafbg, boxrule=0.4pt, arc=2pt, boxsep=0pt, left=1.5pt, right=1.5pt, top=1pt, bottom=1pt]{#1}}}

\newif\ifshownotes
\shownotesfalse 

\makeatletter
\newcommand{\buglink}[3]{\if@ACM@anonymous #2\else\href{#3}{#1}\fi}
\newcommand{\anonnote}[1]{\if@ACM@anonymous #1\fi}
\makeatother

\newcommand{\rqanswer}[1]{\par\vspace{6pt}\noindent\textbf{Answer:} \emph{#1}\par}

\begin{document}
\title{Typed Template Fuzzing}
\author{Lu Maltsis}
\affiliation{%
  \institution{University of Manchester}
  \city{Manchester}
  \country{United Kingdom}}
\email{lu.maltsis@postgrad.manchester.ac.uk}
\author{Dominik Winterer}
\affiliation{%
  \institution{University of Manchester}
  \city{Manchester}
  \country{United Kingdom}}
\email{dominik.winterer@manchester.ac.uk}
\begin{abstract}
    Fuzzing software with complex inputs has been a longstanding challenge.
Traditionally, fuzzers either programmatically encode language constraints, requiring significant
engineering effort, or rely on users to declare grammars and constraints.
Recent LLM-based fuzzers address these drawbacks but introduce new
issues. Some use LLMs for test generation, sacrificing
throughput and producing invalid inputs when the LLM makes mistakes. Others rely
on language-specific parsers, static analysis, and AST manipulation libraries.

We introduce \approachname{} (\approachnameShorthand{}), a lightweight LLM-based
fuzzing methodology. The key idea is to use LLMs in
a preprocessing step to transform seeds into typed templates. The typed templates
then enable semantically aware, mutation-based fuzzing without
language-specific parsers, type checkers, or further LLM queries during fuzzing.
By exploiting the correspondence between seeds and templates, \approachnameShorthand{} can moreover repair
LLM mistakes. We have implemented \approachnameShorthand{} in our tool \toolname{},
which we evaluate across \numInputLanguages{} input languages, including
C, C++, Go, Rust, SystemVerilog, Scala, and Java, as well as libraries such as PyTorch and Qiskit.
In a six-month fuzzing campaign, we reported \reportedBugs{} bugs,
\unknownBugs{} of which were previously unknown. Perhaps most notably,
on average \validityRate{} of \toolname{}'s mutants are valid, coming 
close to the validity rates of domain-specific fuzzers.
Our evaluation further shows \toolname{}'s benefits over the state of the art.

\end{abstract}

\maketitle

\section{Introduction}
Fuzzing is an automated testing technique for generating high input volumes 
to trigger unexpected behavior. Compilers, interpreters, constraint solvers, and 
libraries feature complex input languages, making it challenging to 
generate \emph{semantically valid} input to exercise deep program behavior. 

Traditionally, two families of fuzzers address this challenge.
Domain-specific fuzzers~\cite{yang-etal-pldi11, livinskii-etal-oopsla20} 
programmatically encode input language constraints. 
While often producing valid input at         
a near-perfect rate, they require significant engineering effort to build and 
cannot be easily adapted to evolving languages. Domain-agnostic fuzzers include  
AFL-based~\cite{zalewski-web16, fioraldi-etal-woot20},
grammar-based~\cite{purdom-bit72, hodovan-etal-atest18, wang-etal-icse19},
and language-based fuzzers~\cite{steinhofel-etal-fse22}. These approaches
either achieve low validity rates or require users to declare grammars and   
language constraints.    

Recently, researchers have been using LLMs to address the shortcomings                
of the two fuzzing families (see Table~\ref{tab:fuzzer_comparison} for a taxonomy).     
One way is to use LLMs for test generation in the fuzzing loop~\cite{xia-etal-icse24, yang-etal-oopsla24}. 
While such an approach incurs low user effort, the drawback is that almost   
every mutant relies on a new LLM query, making fuzzing slow and costly. Moreover,          
LLM mistakes cannot be corrected, resulting in a high proportion of invalid mutants.  
LegoFuzz~\cite{ni-etal-oopsla25} uses LLMs to produce a corpus of self-contained 
functions for mutating C programs. While LegoFuzz frees the fuzzing loop from LLM calls, 
it needs C-specific AST manipulation tooling and cannot repair LLM mistakes.         
MetaMut and Once4All~\cite{ou-etal-asplos25, sun-etal-asplos26} use LLMs
to generate code. Both feature repair mechanisms but remain domain-specific    
as the LLM-generated code relies on language-specific tooling
(e.g., AST manipulation, parsing). Most recently, GenToo~\cite{vikram-ase26},
an agentic approach for writing fuzzers for Java libraries, was proposed. While GenToo                   
has an LLM-free fuzzing loop and repair for LLM mistakes, it is domain-specific,
i.e., it relies on Java-specific static analysis.

\paragraph{\textbf{\approachname{}}}
We propose \approachname{} (\approachnameShorthand{}), a lightweight LLM-based    
fuzzing approach. 
Its key idea is to use an LLM to create
typed templates from a set of initial program seeds once, upfront, in a preprocessing step.
The resulting typed templates contain annotations with type and scope information
but are otherwise identical to the original seed programs. This correspondence     
allows us to repair LLM mistakes using a simple repair procedure.         
Given the typed templates, \approachnameShorthand{} chooses an expression at random and replaces it with
a newly generated one of the same type. \approachnameShorthand{} has several distinctive advantages. 
First, \approachnameShorthand{} does not require domain-specific tooling such as a parser or type checker. 
Second, it does not need repeated LLM calls in the fuzzing loop or 
LLM-based iterative refinement. 
We realize \approachname{} in a practical tool \toolname{}, which we have
extensively evaluated across \numInputLanguages{} input languages.
Our results reveal that \toolname{}'s mutated test inputs are valid in
\validityRate{} of cases on average, nearly matching purpose-built domain-specific 
fuzzers and outperforming or matching the previous state of the art in LLM-based 
fuzzing. We further conducted a bug-hunting campaign of \lengthOfBugHunting{} 
months on \numTargets{} targets across \numInputLanguages{}
languages, including \allTargets{}. We found \unknownBugs{} new bugs, of which \fixedBugs{}
have already been fixed by project maintainers. 

The source code of \toolname{}, seed files, prompts, experimental data, and evaluation scripts
are available in our replication package (see \hyperref[sec:data-availability]{Data Availability}).

\begin{table}[!t]
    \centering
    \footnotesize
    \renewcommand{\arraystretch}{1.4}
    \setlength{\tabcolsep}{4pt}
    \renewcommand{\cmark}{\textcolor{green!60!black}{\ding{52}}}
    \renewcommand{\xmark}{\textcolor{red!80!black}{\ding{56}}}
    \caption{Taxonomy of LLM-based fuzzers targeting complex input formats. 
    Existing approaches either rely on language-specific tooling, LLM input generation in the fuzzing loop 
    or cannot repair LLM mistakes.   
    \toolname{} uses the LLM to annotate seed files in preprocessing, requiring no language-specific tooling and no LLM calls in the fuzzing loop, and can repair LLM mistakes.}
    \label{tab:fuzzer_comparison}
    \begin{tabular}{lcccl}
        \toprule
        \textbf{LLM-Based Fuzzer} & \textbf{\begin{tabular}[c]{@{}c@{}}Language-\\Agnostic\end{tabular}} & \textbf{\begin{tabular}[c]{@{}c@{}}LLM-Free\\Fuzzing Loop\end{tabular}} & \textbf{\begin{tabular}[c]{@{}c@{}}Repairs\\LLM Errors\end{tabular}} & \textbf{LLM Task} \\
        \midrule
        Fuzz4All~\cite{xia-etal-icse24}, WhiteFox~\cite{yang-etal-oopsla24}
        & \cmark
        & \xmark
        & \xmark
        & Test input generation\\
        LegoFuzz~\cite{ni-etal-oopsla25} & \xmark & \cmark & \xmark & Code fragment generation\\
        MetaMut~\cite{ou-etal-asplos25}, Once4All~\cite{sun-etal-asplos26}
        & \xmark
        & \cmark
        & \cmark
        & Write fuzzing code\\
        GenToo~\cite{vikram-ase26} & \xmark & \cmark & \cmark & Generator synthesis\\
        \textbf{\toolname{}} (ours)  & \cmark & \cmark & \cmark & \textbf{Annotate seed files} \\
        \bottomrule
    \end{tabular}
\end{table}

\paragraph*{\textbf{Main Contributions}} In sum, we make the following main contributions:  
\begin{itemize}[itemsep=2pt]
    \item We propose \approachname{}, a lightweight LLM-based fuzzing methodology.  
    The key idea is to use an LLM upfront in a preprocessing step to transform seeds into   
    templates that are then used for fuzzing. \approachnameShorthand{} is language-agnostic, has an LLM-free         
    fuzzing loop, and is able to repair LLM mistakes;   
    \item Based on \approachname{}, we implemented an effective tool \toolname{} with
    which we conducted several in-depth evaluations to assess its effectiveness.
    Perhaps most notably, we observe that \toolname{} produces valid inputs
    at a rate of \validityRate{} on average across all languages, which is 
    within a few percentage points of domain-specific fuzzers;  
    \item We conducted a bug-hunting campaign on \numInputLanguages{} input
    languages and \numTargets{} targets during which we reported \reportedBugs{} new bugs,
    out of which \confirmedBugs{} bugs were confirmed and \fixedBugs{} bugs were
    fixed.
\end{itemize}

\section{Illustrative Example}
\label{sec:example}
This section illustrates \approachname{} on an example (Fig.~\ref{fig:templating}). 
The approach has two stages: preprocessing and fuzzing. In preprocessing, we generate 
a typed template and potentially repair it. In fuzzing, we then use the repaired 
typed template to generate vast input volumes by mutation and forward these inputs 
to the system under test to capture unexpected behavior. 

\begin{figure}[H]
\captionsetup[subfigure]{justification=centering,singlelinecheck=false}

\begin{subfigure}[t]{0.31\columnwidth}
\begin{lstlisting}[escapechar=!, basicstyle=\scriptsize\ttfamily, columns=fullflexible, basewidth={0.45em,0.45em}, keepspaces=true, breaklines=false, xleftmargin=0pt, aboveskip=0pt]
func main() {
    i := 8
    output := make([]int, i, i * 2)
    fmt.Println(i, output)
}
\end{lstlisting}

\caption{}\label{fig:seed}
\end{subfigure}%
\hfill
\begin{subfigure}[t]{0.68\columnwidth}
\begin{lstlisting}[escapechar=!, basicstyle=\scriptsize\ttfamily, columns=fullflexible, basewidth={0.45em,0.45em}, keepspaces=true, breaklines=false, xleftmargin=0pt, aboveskip=0pt]
func main() {
    i := !\markA{<|8:int|>}!
    output := make([]int, !\markA{<|i:int:\{i\}|>}!, !\markB{<|\markA{<|i:int:\{i\}|>} * \markA{<|2:int:\{i\}|>}:int|>}!)!\markE{)}!
    !\markE{format}!.Println(!\markA{<|i:int:\{i,output\}|>}!, !\markA{<|output:[]int:\{i,output\}|>}!)
}
\end{lstlisting}

\caption{}\label{fig:llmraw}
\end{subfigure}
\par\vspace{0.4em}
\begin{subfigure}[t]{0.68\columnwidth}
\begin{lstlisting}[escapechar=!, basicstyle=\scriptsize\ttfamily, columns=fullflexible, keepspaces=true, breaklines=false, aboveskip=0pt]
func main() {
    i := !\markA{<|8:int|>}!
    output := make([]int, !\markA{<|i:int:\{i\}|>}!, !\markB{<|\markA{<|i:int:\{i\}|>} * \markA{<|2:int:\{i\}|>}:int|>}!)
    !\markR{fmt}!.Println(!\markA{<|i:int:\{i,output\}|>}!, !\markA{<|output:[]int:\{i,output\}|>}!)
}
\end{lstlisting}

\caption{}\label{fig:templated}
\end{subfigure}
\par\vspace{0.4em}
\begin{subfigure}[t]{0.29\columnwidth}
\begin{lstlisting}[escapechar=!, basicstyle=\scriptsize\ttfamily, columns=fullflexible, keepspaces=true, breaklines=false, aboveskip=0pt]

func main() {
    i := 8
    output := make([]int, i, !\markS{i}!)
    fmt.Println(i, output)
}
\end{lstlisting}

\caption{}\label{fig:m1}
\end{subfigure}%
\hspace{8em}
\begin{subfigure}[t]{0.32\columnwidth}
\begin{lstlisting}[escapechar=!, basicstyle=\scriptsize\ttfamily, columns=fullflexible, keepspaces=true, breaklines=false, aboveskip=0pt]
!{\color{black!40}// \href{https://github.com/golang/go/issues/77919}{golang/go\#77919}}!
func main() {
    i := 8
    output := make([]int, !\markS{8 * 8}!, i)
    fmt.Println(i, output)
}
\end{lstlisting}

\caption{}\label{fig:m2}
\end{subfigure}
\caption{Typed Template Fuzzing illustrated. (a) Initial Go program. (b) LLM-created typed template. Expressions boxed, mistakes are highlighted in red. (c) Repaired typed template. (d+e) Two consecutive mutants are generated in the fuzzing loop. The second mutant triggers an internal compiler error in gc, the default Go compiler.\anonnote{ The bug report number is masked to preserve anonymity.}                   
\label{fig:templating}}   
\end{figure}

\begin{description}[itemsep=0.5em]
    \item[] \textbf{Preprocessing}: Given an initial Go program (Fig.~\ref{fig:seed}),
    we first query an LLM to transform the program into a typed template (Fig.~\ref{fig:llmraw}).  
    In this process, expressions are annotated with
    their default value, type, and scoping information. E.g., \markA{\ttfamily<|i:int:\{i,output\}|>}
    indicates a default value of "\texttt{i}", type "\texttt{int}", and both  
    variables "\texttt{i}" and "\texttt{output}" in scope.   
    Subsequently, we correct potential mistakes by the LLM using a 
    repair procedure. The repair procedure realizes a character-by-character comparison   
    between the initial program and the generated template.     
    In our example, the procedure captures an excess parenthesis and          
    a wrong identifier. This results in a corrected template (Fig.~\ref{fig:templated}).    
    A further preprocessing step learns operators such as \markA{\ttfamily<int> * <int>}
    from annotated nested expressions.   
    \item[] \textbf{Fuzzing loop}: Having a typed template, the fuzzing loop 
    iteratively generates new programs by mutation to stress-test a system under 
    test, i.e., the Go compiler in our case. One possible mutation is substituting 
    the expression "\texttt{i * 2}" with the variable "\texttt{i}". The resulting program 
    (Fig.~\ref{fig:m1}) is semantically valid and successfully compiled by the Go 
    compiler. Generating fresh expressions is enabled by operators.         
    For example, we use the derived operator \markA{\ttfamily<int> * <int>}
    to create the new expression "\texttt{8 * 8}". 
    In our example, mutating the expression "\texttt{i}" creates another mutant (Fig.~\ref{fig:m2}), 
    triggering a previously undetected internal compiler error (ICE) in the Go compiler gc. We reported
    the ICE on the GitHub issue tracker and the underlying issue was promptly fixed by the developers. 

\end{description}

\section{\approachname{}}
\label{sec:approach}
This section first introduces basic fuzzing terminology, then formalizes typed templates, and finally defines mutation and generation as well as Typed Template Fuzzing. 

\paragraph{\textbf{Fuzzing terminology}} A \emph{system under test} (SUT) is software 
that takes a program as input and exhibits behavior such as exit codes, crashes, 
\etc{} A \emph{fuzzer} is a program that generates large volumes of inputs
for an SUT. We use the terms inputs and programs interchangeably. A program is \emph{valid}
with respect to an SUT if it complies with the constraints of the SUT's input language.
In practice, we consider a program semantically valid if the SUT accepts it, i.e., terminates
with exit code zero. \emph{Seeds} denote a set of valid initial programs given
to a fuzzer. \emph{Mutants} denote programs generated by a fuzzer. 
The \emph{fuzzing loop} describes the iterative process of turning seed programs 
into mutants. The goal of a fuzzer in our context is to generate        
valid mutants at the highest possible rate to exercise deep behavior in the     
SUTs.

\subsection{Typed templates}
The key idea behind typed templates is to process programs as text with annotated substrings. 
Hence we begin with basic definitions on strings. A \emph{string} $w$ is a finite sequence of characters.
We write $|w|$ for its length. A \emph{span} $(i,j)$ over $w$, with $0 \le i \le j \le |w|$, 
denotes the position of the substring $w[i\;{:}\;j]$ consisting of the characters 
at positions $i$ to $j-1$. We write $w_1 \cdot w_2$ for the concatenation of strings 
$w_1$ and $w_2$. We assume a fixed alphabet $\Sigma$. A program $\program$ is a 
string over $\Sigma$. 

\begin{definition}[Typed template]
\label{def:typed_template}
A \emph{typed template} $\ttemplate{\program}$ created from a program $\program$ is a string over $\Sigma$ that contains annotated substrings, called \emph{terms}.
Each term $e$ has
\begin{itemize}
  \item $\body{e}$, its \emph{body},
  \item $\sort{e} \in \{ \mathrm{Var}, \mathrm{Const}, \mathrm{Expr}\}$, its \emph{sort},
  \item $\type{e}$, its \emph{type}, which is a string, and
  \item $\live{e}$, its \emph{live variables}, i.e., the variables in scope at $e$, as a set of typed variables $x{:}T$, each with variable name $x$ and type $T$.
\end{itemize}
Any two terms are either disjoint or one contains the other. Terms of sort $\mathrm{Expr}$ can contain other terms.
The top-level terms contained in a term $e$ are called \emph{operands} $\operands(e)$.
\end{definition}

We represent a term $e$ by \texttt{<|}$\body{e}$\texttt{:}$\sort{e}$\texttt{:}$\type{e}$\texttt{:\{}$\live{e}$\texttt{\}|>}.
We assume that the delimiters 
in this representation allow for a unique matching of terms. For compactness,
we omit the sort, when evident from the body, and the types of live variables.
We also omit the live variables entirely when evident from the context.           
We illustrate this notation in the following Java program. 

\begin{center}
\begin{minipage}[t]{76pt}
\begin{lstlisting}[escapechar=!, basicstyle=\footnotesize\ttfamily, columns=fullflexible, keepspaces=true, breaklines=false,
aboveskip=0pt, belowskip=0pt]
class A {
  int foo(int x) {
      return x + 4;
  }
}
\end{lstlisting}
\end{minipage}%
\hspace{3em}%
\begin{minipage}[t]{175pt}
\begin{lstlisting}[escapechar=!, basicstyle=\footnotesize\ttfamily, columns=fullflexible, keepspaces=true, breaklines=false,
aboveskip=0pt, belowskip=0pt]
class A {
  int foo(int x) {
      return !\markB{<|\markA{<|x:int:\{x\}|>} + \markA{<|4:int:\{x\}|>}:int|>}!;
  }
}
\end{lstlisting}
\end{minipage}
\vspace{0.5em}
\end{center}
In the typed template on the right-hand side, the variable \markAt{<|x:int:\{x\}|>}, the constant \markAt{<|4:int:\{x\}|>} and the expression \markBt{<|\markAt{<|x:int:\{x\}|>} + \markAt{<|4:int:\{x\}|>}:int|>} are annotated, and the expression contains the other two as operands.
Turning templates back into programs, we define: for a string $w$, its \emph{instantiation} $\content{w}$ denotes the string obtained by recursively replacing every term in $w$ by its body.
For example,
\[
  \content{\mbox{\markBt{<|\markAt{<|x:int:\{x\}|>} + \markAt{<|4:int:\{x\}|>}:int|>}}} = \texttt{x + 4}.
\]
Applied to the whole typed template, $\content{\cdot}$ yields the original program on the left.
Since typed templates may be created through error-prone processes, including LLMs or human annotation, we define quality criteria that ensure their usefulness.
\needspace{10\baselineskip}
\begin{definition}[Template Quality Criteria]
\label{def:quality}
A typed template $\ttemplate{\program}$ is
\begin{enumerate}
  \item \emph{grounded} if $\content{\ttemplate{\program}} = \program$,
  \item \emph{variable-complete} if every term of $\ttemplate{\program}$ is variable-complete, where a term $e$ is \emph{variable-complete} if every variable inside $e$ is annotated as a $\mathrm{Var}$ term, and
  \item \emph{annotation-sound} if every term of $\ttemplate{\program}$ is annotation-sound, where a term $e$ is \emph{annotation-sound} if $e$ and every term inside it annotate an expression of $\content{\ttemplate{\program}}$ and match its \emph{actual} sort, type and live variables in that program.
\end{enumerate}
\end{definition}
Groundedness ensures that the template generation process preserves the text of the original program, so that instantiating the template yields the original, valid seed.
Variable-completeness intuitively means that we know all variables that a term depends on. In turn, this allows mutations to respect variable scoping.
Annotation-soundness ensures that the annotated types and live variables reflect the semantics of $\program{}$.
As an example, consider a TypeScript program along with a template that satisfies all three properties and one
that violates them all (Fig.~\ref{fig:wellformed}).
The program first defines the variable \texttt{x} as the number \texttt{3} and then the variable \texttt{y} as \texttt{x * 2}.
The first template (Fig.~\ref{fig:wf-wellformed}) is grounded, since its instantiation reproduces the original program.
By contrast, the second template (Fig.~\ref{fig:wf-illformed}) contains a typo in the first \texttt{let} statement and is therefore not grounded.
As \texttt{x} inside \texttt{x * 2} is annotated in the first template, it is variable-complete, whereas the second template lacks this annotation and is not.
Additionally, \texttt{2} is mistakenly annotated as a \texttt{string} in the second template instead of \texttt{number} as in the annotation-sound first template.

\newsavebox{\wfbox}
\begin{figure}[t]
\centering
\begin{subfigure}[t]{0.115\columnwidth}
\centering
\begin{lrbox}{\wfbox}\begin{varwidth}[t]{\textwidth}
\begin{lstlisting}[escapechar=!, basicstyle=\tiny\ttfamily, columns=fullflexible, keepspaces=true, breaklines=false, aboveskip=0pt]
let x = 3;
let y = x * 2;
\end{lstlisting}
\end{varwidth}\end{lrbox}\scalebox{1.08}{\usebox{\wfbox}}
\caption{}\label{fig:wf-original}
\end{subfigure}%
\hfill
\begin{subfigure}[t]{0.47\columnwidth}
\centering
\begin{lrbox}{\wfbox}\begin{varwidth}[t]{\textwidth}
\begin{lstlisting}[escapechar=!, basicstyle=\tiny\ttfamily, columns=fullflexible, keepspaces=true, breaklines=false, aboveskip=0pt]
let x = !\markA{<|3:number:\{\}|>}!;
let y = !\markB{<|\markA{<|x:number:\{x\}|>} * \markA{<|2:number:\{x\}|>}:number|>}!;
\end{lstlisting}
\end{varwidth}\end{lrbox}\scalebox{1.08}{\usebox{\wfbox}}
\caption{}\label{fig:wf-wellformed}
\end{subfigure}%
\hfill
\begin{subfigure}[t]{0.345\columnwidth}
\centering
\begin{lrbox}{\wfbox}\begin{varwidth}[t]{\textwidth}
\begin{lstlisting}[escapechar=!, basicstyle=\tiny\ttfamily, columns=fullflexible, keepspaces=true, breaklines=false, aboveskip=0pt]
!\markE{lte}! x = !\markA{<|3:number:\{\}|>}!;
let y = !\markB{<|\markE{x} * \markA{<|2:\markE{string}:\{x\}|>}:number|>}!;
\end{lstlisting}
\end{varwidth}\end{lrbox}\scalebox{1.08}{\usebox{\wfbox}}
\caption{}\label{fig:wf-illformed}
\end{subfigure}
\caption{(a) TypeScript program.
(b) Grounded, variable-complete and annotation-sound template.
(c) Template that is neither grounded (\texttt{lte}), nor variable-complete (\texttt{x} not annotated), nor annotation-sound (\texttt{string}).}
\label{fig:wellformed}
\end{figure}

\subsection{Mutation and Generation with Typed Templates}
We next define mutation and generation with typed templates.
We begin with the definition of substitution, then define term generation and Typed Template Fuzzing.
\noindent
\begin{definition}[Substitution]
\label{def:substitution}
The \emph{substitution} of a term $e$ by another term $e'$ is 
$$\subst{w}{e'}{e} = w[0\;{:}\;i] \cdot e' \cdot w[j\;{:}\;|w|]$$
where $e$ occurs in a string $w$ at span $(i,j)$. 
\end{definition}

We emphasize that $e$ refers to a single occurrence at span $(i,j)$.
We further assume that the spans of all other terms in $w$ are updated accordingly. 
A necessary ingredient for creating valid mutants in the fuzzing loop 
is to ensure that terms do not refer to undefined variables when substituted.
To that end, we define a notion of free variables.
For a term $e$, the \emph{free variables} of $e$ are defined by
\[
  \freevars{e} \;=\;
  \begin{cases}
    \{\content{e}{:}\type{e}\} & \sort{e} = \mathrm{Var} \\
    \emptyset & \sort{e} = \mathrm{Const} \\
    \bigcup_{g \,\in\, \operands(e)} \freevars{g} & \sort{e} = \mathrm{Expr}
  \end{cases}
\]
As templates do not record binders in operands, $\freevars{e}$ over-approximates 
the free variables of a variable-complete term $e$.
It remains to define when a term can replace another.
A term $e'$ is \emph{admissible} at term $e$ if $\type{e'} = \type{e}$ and $\freevars{e'} \subseteq \live{e}$.
Admissibility constrains mutations based on typing and variable scoping.

\begin{definition}[Term Generation]
    Let $g$ be a term with $\sort{g} = \mathrm{Expr}$ and let $\operands(g) = (g_1, \ldots, g_m)$.
    We call $g$ an operator. Given $r_i$ with $\type{r_i} = \type{g_i}$ for all $1 \leq i \leq m$, 
    we define a replacement term $r = g[r_1 / g_1][r_2 / g_2] \ldots [r_m / g_m]$. 
    We call the terms $r_i$ chosen terms.  
    \label{def:generative_term}
\end{definition}

\begin{definition}[\approachname{}]
    Let $e$ be a term in template $\ttemplate{\program}$ and let $r$ be a replacement term. 
    Require further that $\type{g} = \type{e}$ and $\freevars{r_i} \subseteq \live{e}$ for all $1 \leq i \leq m$.
    We call $\subst{\ttemplate{\program}}{r}{e}$ a \emph{mutated template} and its instantiation $\content{\subst{\ttemplate{\program}}{r}{e}}$ a \emph{mutant}.
    \label{def:generative_mutation}
\end{definition}
In the mutated template, we set the live variables of $r$ and of every term inside $r$
to $\live{e}$. The mutated template is again a typed template, since substitution
preserves the nesting of terms, and can thus itself be mutated.
For instance, consider the 
following typed template: 
\begin{center}
\begin{minipage}[c]{200pt}
\begin{lstlisting}[escapechar=!, basicstyle=\footnotesize\ttfamily, columns=fullflexible, keepspaces=true, breaklines=false,
aboveskip=0pt, belowskip=0pt]
let x = !\markA{<|3:number:\{\}|>}!;
let y = !\markB{<|\markA{<|x:number:\{x\}|>} * \markA{<|2:number:\{x\}|>}:number|>}!;
\end{lstlisting}
\end{minipage}
\vspace{0.5em}
\end{center}
Let $e$ be the constant \markAt{<|3:number:\{\}|>} in the first line, where $\live{e} = \emptyset$ since \texttt{x} is not yet defined.
As the operator, we take the multiplication $g$ from the second line, with operands $g_1$ = \markAt{<|x:number|>} and $g_2$ = \markAt{<|2:number|>}.
As chosen terms, we take the constants $r_1$ = \markAt{<|2:number|>} and $r_2$ = \markAt{<|3:number|>}.
Both conditions of Def.~\ref{def:generative_mutation} hold: $\type{g} = \type{e} = \texttt{number}$, and $\freevars{r_1} = \freevars{r_2} = \emptyset \subseteq \live{e}$.
Substituting the replacement term $r$ for $e$ yields the mutated template
\begin{center}
\begin{minipage}[c]{200pt}
\begin{lstlisting}[escapechar=!, basicstyle=\footnotesize\ttfamily, columns=fullflexible, keepspaces=true, breaklines=false,
aboveskip=0pt, belowskip=0pt]
let x = !\markB{<|\markA{<|2:number:\{\}|>} * \markA{<|3:number:\{\}|>}:number:\{\}|>}!;
let y = !\markB{<|\markA{<|x:number:\{x\}|>} * \markA{<|2:number:\{x\}|>}:number|>}!;
\end{lstlisting}
\end{minipage}
\vspace{0.5em}
\end{center}
and thus the mutant TypeScript program 
\begin{center}
\begin{minipage}[c]{80pt}
\begin{lstlisting}[basicstyle=\footnotesize\ttfamily, columns=fullflexible, keepspaces=true, breaklines=false,
aboveskip=0pt, belowskip=0pt]
let x = 2 * 3;
let y = x * 2;
\end{lstlisting}
\end{minipage}
\vspace{0.5em}
\end{center}
By contrast, choosing the variable \markAt{<|x:number|>} as a chosen term violates the second condition, since $\freevars{\texttt{x}} = \{\texttt{x}{:}\texttt{number}\} \not\subseteq \live{e}$.
We also note that replacing a term directly by a chosen term, i.e., non-generative mutation, can be realized by an identity operator, whose body consists of a single operand.
In general, operator $g$ and chosen terms $r_i$ may stem from any template of the same language, which lets terms transfer across programs.

\section{\toolname{}'s implementation}
\label{sec:implementation}
This section presents \toolname{}, our realization of \approachname{}. We first           
present a brief approach overview, then address practical challenges 
along with their mitigation strategies. 

\begin{figure*}[t]
\input{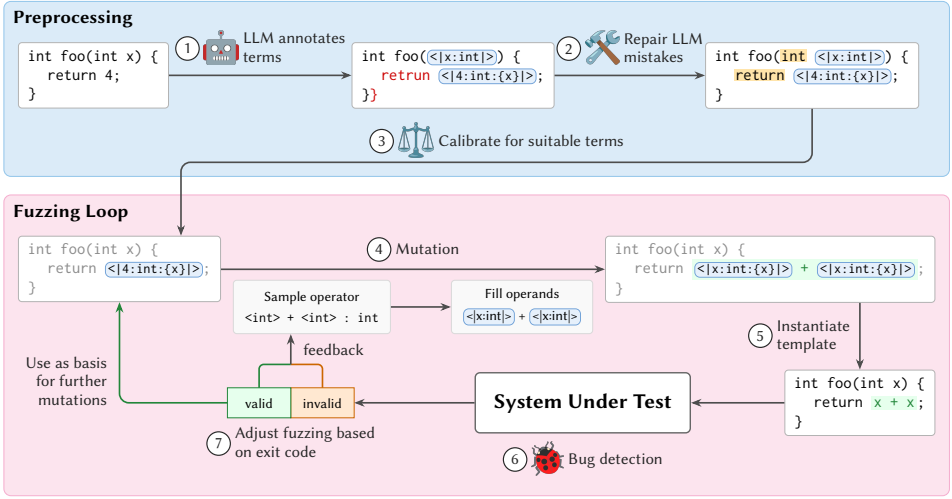}
\caption{An overview of the pipeline of \toolname{}. Seed files are
first preprocessed into templates and then repeatedly mutated.
Mutated templates of valid mutants are reused for further mutation, and the SUT's exit code guides operator selection.}
\label{fig:overview}
\end{figure*}

\begin{figure}[!t]
\input{includes/prompt}
\caption{Template generation prompts. Pass~1 tags variables, constants, and operator symbols, and Pass~2 groups these tags into nested expressions; \texttt{[...]} marks omitted parts of the prompts.}
\label{fig:prompts}
\end{figure}

\subsection{\toolname{} overview}
\toolname{} consists of two stages: preprocessing and the fuzzing loop 
(see Fig.~\ref{fig:overview}). The pipeline takes a set of seeds; however, for simplicity, we illustrate it on a single seed. During preprocessing, an LLM annotates 
the terms in a seed program with their type and live variables, producing a typed 
template (Step~1). Templates augment programs with inline annotations, 
e.g.\ \markAt{<|4:int:\{x\}|>} represents the term \texttt{4} of type 
\texttt{int} and live variable \texttt{x}. The typed template is then repaired 
to fix LLM mistakes (Step~2) and calibrated to keep only terms suitable for mutation (Step~3).
The mutation loop is \toolname{}'s fuzzing loop: it repeatedly applies 
\approachname{} (Def.~\ref{def:generative_mutation}). Each iteration replaces a 
term with a newly generated one (Step~4), instantiates the result (Step~5), and 
runs it on the system under test to detect potential bugs (Step~6).
Based on the SUT's exit code, the mutated templates of valid mutants are kept for 
further mutation and operator selection is adjusted (Step~7).
Overall, \toolname{} is realized in 2,047 lines of Python code.  

\paragraph{\textbf{Template generation}}
Typed templates are created by prompting an LLM with a language-specific task 
description and a few examples~\cite{brown-etal-neurips20}. The process operates 
in two passes (Fig.~\ref{fig:prompts}).
First, it annotates variables, constants and operator symbols (e.g.\ \texttt{+}, \texttt{*}).
In a second pass, it annotates nested expressions, with those containing an operator symbol serving as operators (Def.~\ref{def:generative_term}).
Since the LLM needs to re-emit the program 
after each pass, the text between terms diverges from the original.

\paragraph{\textbf{Practical challenges}}
A major challenge for \toolname{} is ungrounded templates. This is because an 
ungrounded template implies that almost all mutants instantiated are  
invalid. We address this by a repair procedure 
(Section~\ref{sec:ungrounded-templates}). Another challenge is variable-incompleteness 
in chosen terms. Mutation with variable-incomplete chosen terms 
leads to invalid mutants in the fuzzing loop. We mitigate the issue by 
sampling (Section~\ref{sec:chosen-terms}). An orthogonal challenge is that the     
LLM may identify terms that do not allow repeated substitution (e.g., terms within 
keywords). We mitigate the issue by a calibration procedure during preprocessing    
(Section~\ref{sec:unsubstitutable}).

\begin{algorithm}[t]
\caption{Repair Procedure}
\small\linespread{1.1}\selectfont
\begin{algorithmic}[1]
\Procedure{Repair}{$\program, \ttemplate{\program}$}
    \State $\textrm{terms} \gets$ outermost terms of $\ttemplate{\program}$, left to right
        \Comment{only the terms are taken from the LLM output}
    \State $\textrm{pos} \gets 0$
        \Comment{cursor into $\program$}
    \State $\textrm{result} \gets \varepsilon$
    \ForAll{$e \in \textrm{terms}$}
        \State $i \gets \program\textrm{.find}(\content{e}, \textrm{start\_at}{=}\textrm{pos})$\label{alg:repair:find}
        \State $\textrm{result} \gets \textrm{result} \cdot \program[\textrm{pos}\;{:}\;i]$\label{alg:repair:gap} \Comment{reconstruct text from $\program$}
        \State $\textrm{result} \gets \textrm{result} \cdot e$\label{alg:repair:term}
        \State $\textrm{pos} \gets i + |\content{e}|$
    \EndFor
    \State $\textrm{result} \gets \textrm{result} \cdot \program[\textrm{pos}\;{:}\;|\program|]$\label{alg:repair:tail}
    \State \Return $\textrm{result}$
\EndProcedure
\end{algorithmic}
\label{alg:repair}
\end{algorithm}

\subsection{Repairing ungrounded templates}
\label{sec:ungrounded-templates}
\providecolor{tagleaf}{RGB}{110,160,215}
\providecolor{tagleafbg}{RGB}{230,240,251}
\definecolor{origseg}{RGB}{255,228,170}
\definecolor{origsegdark}{RGB}{200,140,20}
\definecolor{tagleafdark}{RGB}{50,100,170}
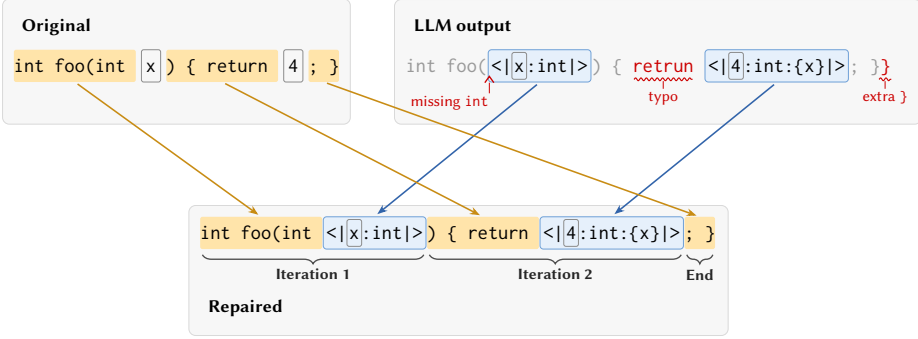
\begin{figure}[t]
\centering
\newcommand*{\rsp}{\phantom{a}}
\scalebox{0.85}{%
\begin{tikzpicture}[
  rb/.style={font=\ttfamily\small, inner xsep=0pt, inner ysep=2pt, outer sep=0pt, text height=1.8ex, text depth=.6ex, anchor=west},
  seg/.style={rb, fill=origseg, rounded corners=1pt},
  slot/.style={rb, inner xsep=2pt, draw=black!45, line width=.4pt, rounded corners=1pt},
  tval/.style={rb, inner xsep=1pt, inner ysep=.8pt, draw=black!45, line width=.4pt, rounded corners=1pt},
  termbox/.style={fill=tagleafbg, draw=tagleaf, line width=.4pt, rounded corners=1.5pt, inner xsep=1.5pt, inner ysep=0pt},
  llm/.style={rb, text=black!40},
  err/.style={rb, text=red!80!black},
  mv/.style={-{Stealth[length=4.5pt]}, line width=.7pt, shorten >=1.5pt, shorten <=1.5pt},
  num/.style={circle, fill=white, draw, line width=.5pt, inner sep=0pt, minimum size=10pt, font=\sffamily\scriptsize},
  numend/.style={rectangle, rounded corners=5pt, fill=white, draw, line width=.5pt, inner xsep=2pt, inner ysep=0pt, minimum height=10pt, font=\sffamily\scriptsize},
  title/.style={anchor=south west, font=\sffamily\footnotesize\bfseries},
  errlab/.style={anchor=north, font=\sffamily\scriptsize, text=red!80!black, inner sep=1pt},
  squiggle/.style={red!80!black, line width=.5pt, decorate, decoration={snake, amplitude=.6pt, segment length=2.4pt}},
  errtick/.style={red!80!black, line width=.4pt},
  panel/.style={fill=black!3, draw=black!15, line width=.4pt, rounded corners=3pt, inner sep=5pt},
  itbrace/.style={decorate, decoration={brace, mirror, amplitude=4pt, raise=2.5pt}, line width=.6pt, black!70},
  itlab/.style={midway, below=7.5pt, align=center, font=\sffamily\scriptsize, text=black!85, inner sep=1pt},
]
\node[seg]  (o1) at (0,0)     {int foo(int\rsp{}};
\node[slot] (o2) at ([xshift=2.5pt]o1.east) {x};
\node[seg]  (o3) at ([xshift=2.5pt]o2.east) {)\rsp{}\{\rsp{}return\rsp{}};
\node[slot] (o4) at ([xshift=2.5pt]o3.east) {4};
\node[seg]  (o5) at ([xshift=2.5pt]o4.east) {;\rsp{}\}};
\node[llm]  (l1)  at (6.1,0)              {int foo(};
\node[rb]   (l2a) at ([xshift=2pt]l1.east) {<|};
\node[tval] (l2v) at (l2a.east)            {x};
\node[rb]   (l2b) at (l2v.east)            {:int|>};
\node[llm]  (l3)  at ([xshift=2pt]l2b.east) {)\rsp{}\{\rsp{}};
\node[err]  (l4)  at (l3.east)             {retrun};
\node[llm]  (l5)  at (l4.east)             {\rsp{}};
\node[rb]   (l6a) at ([xshift=2pt]l5.east) {<|};
\node[tval] (l6v) at (l6a.east)            {4};
\node[rb]   (l6b) at (l6v.east)            {:int:\{x\}|>};
\node[llm]  (l7)  at ([xshift=2pt]l6b.east) {;\rsp{}\}};
\node[err]  (l8)  at (l7.east)             {\}};
\node[seg]  (r1)  at (2.9,-2.6)            {int foo(int\rsp{}};
\node[rb]   (r2a) at ([xshift=2pt]r1.east) {<|};
\node[tval] (r2v) at (r2a.east)            {x};
\node[rb]   (r2b) at (r2v.east)            {:int|>};
\node[seg]  (r3)  at ([xshift=2pt]r2b.east) {)\rsp{}\{\rsp{}return\rsp{}};
\node[rb]   (r4a) at ([xshift=2pt]r3.east) {<|};
\node[tval] (r4v) at (r4a.east)            {4};
\node[rb]   (r4b) at (r4v.east)            {:int:\{x\}|>};
\node[seg]  (r5)  at ([xshift=2pt]r4b.east) {;\rsp{}\}};
\node[title] (to) at ([yshift=3pt]o1.north west) {Original};
\node[title] (tl) at ([yshift=3pt]l1.north west) {LLM output};
\coordinate (c1) at ([xshift=-1pt]l2a.south west);
\draw[red!80!black, line width=.6pt] ([xshift=-2pt,yshift=-.3pt]c1) -- ++(2pt,2.8pt) -- ++(2pt,-2.8pt);
\draw[errtick] ([yshift=-.3pt]c1) -- ([yshift=-5pt]c1);
\node[errlab, anchor=north east] (e1) at ([xshift=.8pt,yshift=-5pt]c1) {missing \texttt{int}};
\draw[squiggle] ([xshift=.5pt,yshift=1.2pt]l4.south west) -- ([xshift=-.5pt,yshift=1.2pt]l4.south east);
\draw[errtick] ([yshift=.4pt]l4.south) -- ([yshift=-3pt]l4.south);
\node[errlab] (e2) at ([yshift=-3pt]l4.south) {typo};
\draw[squiggle] ([xshift=-.5pt,yshift=1.2pt]l8.south west) -- ([xshift=.5pt,yshift=1.2pt]l8.south east);
\draw[errtick] ([yshift=.4pt]l8.south) -- ([yshift=-3pt]l8.south);
\node[errlab] (e3) at ([yshift=-3pt]l8.south) {extra \texttt{\}}};
\draw[itbrace] ([xshift=1pt]r1.south west) -- ([xshift=.5pt]r2b.south east) node[itlab] (i1) {\textbf{Iteration 1}};
\draw[itbrace] ([xshift=1pt]r3.south west) -- ([xshift=.5pt]r4b.south east) node[itlab] (i2) {\textbf{Iteration 2}};
\draw[itbrace] ([xshift=1pt]r5.south west) -- ([xshift=-1pt]r5.south east) node[itlab] (i3) {\textbf{End}};
\node[title, anchor=north west] (tr) at ([yshift=-4pt]r1.south west |- i1.south) {Repaired};
\begin{scope}[on background layer]
  \node[panel, fit=(to)(o1)(o5)(o1.west |- e1.south)] {};
  \node[panel, fit=(tl)(l1)(l8)(e1)(e3)] {};
  \node[panel, fit=(tr)(r1)(r5)(i1)(i3)] {};
  \node[termbox, fit=(l2a)(l2b)] (l2) {};
  \node[termbox, fit=(l6a)(l6b)] (l6) {};
  \node[termbox, fit=(r2a)(r2b)] (r2) {};
  \node[termbox, fit=(r4a)(r4b)] (r4) {};
\end{scope}
\draw[mv, origsegdark] (o1.south) -- (r1.north);
\draw[mv, tagleafdark] (l2.south) -- (r2.north);
\draw[mv, origsegdark] (o3.south) -- (r3.north);
\draw[mv, tagleafdark] (l6.south) -- (r4.north);
\draw[mv, origsegdark] (o5.south) -- (r5.north);
\end{tikzpicture}%
}
\caption{Repair procedure on an LLM-generated template. Terms (blue) are taken from the LLM output, the text between them (amber) from the original program. Grey text between terms is LLM-generated and discarded; it contains three errors (red), which repair thereby removes.}
\label{fig:repair-example}
\end{figure}

Since the LLM needs to re-emit the entire program, the text between terms diverges 
from the original, i.e., the template is no longer grounded. Our key observation 
for fixing this is that the LLM output is needed only for terms. 
Everything between them can be retrieved from the original program. Based on this 
insight, we devise a repair procedure (Algorithm~\ref{alg:repair}).      
After each pass, the procedure therefore keeps the terms and re-grounds the text 
between them from the original. Let us consider Fig.~\ref{fig:repair-example}.
Repair maintains a cursor into the original, initially at its start.
In the first iteration, repair searches the original for the text of the 
first term, \texttt{x} (line~\ref{alg:repair:find}), copies the original text 
preceding it into the result (line~\ref{alg:repair:gap}), 
and appends the term (line~\ref{alg:repair:term}); this restores the dropped \texttt{int}.
The cursor now points past \texttt{x}. The second iteration locates \texttt{4} from 
the cursor on and copies the text in between, replacing the misspelled \texttt{retrun}.
Finally, the remaining original text is appended (line~\ref{alg:repair:tail}), which discards the extra curly bracket.
Since all text between terms is taken from the original valid program, all LLM mistakes outside terms are repaired.
Terms not found in the original program are dropped, preserving groundedness.

\subsection{Variable incompleteness in chosen terms}
\label{sec:chosen-terms}
In the fuzzing loop, we pick a template and one of its terms uniformly at random, and generate a new term for it via operators (Def.~\ref{def:generative_term}).
Operands are drawn from the terms of the current seed file and its valid mutants, so each has already appeared in a valid program from this seed.
In contrast, operators are collected from the annotated nested expressions of all templates and used across all files.
Consequently, some operators consistently create invalid mutants.
For example, the following operator, which is not variable-complete (Def.~\ref{def:quality}), triggers scoping errors in any context that does not define \texttt{y}:
\begin{center}
\texttt{\markAt{<|x:int|>} + y}
\end{center}
Other operators depend on context from the original file, such as imports or locally defined functions:
\begin{center}
\texttt{math.sqrt(\markAt{<|x:int|>})} \qquad \texttt{bar(\markAt{<|y:int|>})}
\end{center}

Our key observation is that, unlike terms, operators cannot be probed in isolation: the validity of each generated term relies on both the operator and the operands.
We therefore learn the reliability of each operator gradually during fuzzing rather than testing it ahead of time.
We select operators via Thompson sampling~\cite{thompson-biometrika33}.
Each operator keeps a $\mathrm{Beta}(s+1, f+1)$ distribution, shared across all seed files, where $s$ and $f$ count the valid and invalid mutants it has produced.
To select an operator, we draw one sample from the distribution of each operator of matching type and choose the highest.
The Beta distribution models our current belief about the probability that each operator yields a valid mutant, while the random sampling still allows operators which may have initially been unlucky to be drawn again.

\begin{algorithm}[t]
\caption{Calibration Procedure}
\small\linespread{1.1}\selectfont
\begin{algorithmic}[1]
\Procedure{Calibrate}{$\ttemplate{\program}$}
    \ForAll{terms $e$ of $\ttemplate{\program}$}
        \State $C \gets$ up to $N$ distinct constants $c$ in $\ttemplate{\program}$ with $\type{c} = \type{e}$, $\content{c} \neq \content{e}$\label{alg:calibrate:probes}
        \If{$C = \emptyset$ \textbf{or} some $c \in C$ has $\content{\subst{\ttemplate{\program}}{c}{e}}$ rejected by the SUT}\label{alg:calibrate:check}
            \State strip the annotation of $e$ in $\ttemplate{\program}$\label{alg:calibrate:strip}
                \Comment{$e$ becomes plain text}
        \EndIf
    \EndFor
    \State \Return $\ttemplate{\program}$
\EndProcedure
\end{algorithmic}
\label{alg:calibrate}
\end{algorithm}

\subsection{Non-substitutable terms}
\label{sec:unsubstitutable}
Even in a grounded template, an admissible substitution can yield an invalid program.
Some terms are \emph{non-substitutable}: substituting them can create invalid programs even when the replacement is admissible.
For example, substituting \texttt{o} splits the identifier \texttt{foo}:
\begin{center}
\texttt{f\markAt{<|o:int|>}o(1)} \quad$\longrightarrow$\quad \texttt{f42o(1)}
\end{center}
Another case is l-value terms, which exist in many languages.
Substituting \texttt{x} with any constant expression is rejected:
\begin{center}
\texttt{\markAt{<|x:int|>} = 5} \quad$\longrightarrow$\quad \texttt{3 = 5}
\end{center}

Our key observation is that such failures are largely independent of the replacement.
We therefore probe terms with constants: they contain no free variables, meaning they are admissible based on type alone, and, unlike variables, they can never act as l-values.
Before fuzzing a seed, a lightweight calibration performs this probing (Algorithm~\ref{alg:calibrate}).
It substitutes each term with up to $N$ distinct constants of the same type from the same file (line~\ref{alg:calibrate:probes}).
If any substitution fails, or no such constant exists (line~\ref{alg:calibrate:check}), the term's annotation is stripped, leaving it plain text (line~\ref{alg:calibrate:strip}).
This leaves fewer terms to mutate while preserving groundedness, and raises the validity rate of mutants.
In our implementation we use $N = 2$.

\section{Evaluation}
\label{sec:evaluation}
This section extensively evaluates \toolname{}, our realization of \approachname{}.   
We first present the setup and then present the evaluation based       
on three consecutive research questions.

\subsection{Experimental Setup}
\paragraph{\textbf{Input languages and systems under test}}
We apply \toolname{} to \numInputLanguages{} input languages 
implemented across \numTargets{} systems under test.
We choose these languages as they represent a wide range of the most popular 
statically typed imperative languages (C, C++, C\#, Go, Java, Kotlin, Rust, Solidity, 
TypeScript), dynamic scripting and query languages (JavaScript, Lua, Perl, 
PHP, Python, SQL), functional languages (Erlang, Scala), as well as declarative 
logic, machine-learning and circuit formats (SMT-LIB2, PyTorch, Qiskit, SystemVerilog). 
As the systems under test, we chose up to two compilers for each language in RQ1.     
For RQ2 and RQ3, we narrow down the SUTs 
to one representative and the subset of SUTs supported by the direct competitors, respectively.

\paragraph{\textbf{Seed files \& template generation}}
\toolname{} requires seed files and a template generation prompt. For each language,       
we collected seed files from their regression test suites, public repositories on GitHub   
and other fuzzers for the input language at hand making a total of \TotalSeeds{}
files. We handcrafted a prompt for the Go language. The prompts for all other  
languages were translated from the Go prompt via frontier models such 
as Anthropic's Opus and Sonnet.

\paragraph{\textbf{LLM and hardware}}
For template generation during \toolname{}'s preprocessing, we use Alibaba's LLM Qwen3-Coder-Next 
\cite{cao-etal-arxiv26}. The estimated total cost for conducting this step on all \TotalSeeds{} 
seeds is \$\TotalCost{}, i.e., on average 
\$10.26 per input language.\footnote{The cost estimate is based on \$0.12/\$0.80 per million input/output tokens, as 
listed on \href{https://openrouter.ai/qwen/qwen3-coder-next}{openrouter.ai} [26.08.2026].}
For the fuzzing campaign with \toolname{} (RQ1), we use an AMD EPYC 9654 machine    
equipped with 96 CPU cores running an Ubuntu 24.04. For template generation and comparison 
against other fuzzers requiring LLM inference during fuzzing, we use another         
AMD Ryzen Threadripper 9960X machine with two NVIDIA GeForce RTX 5090 GPUs
running Ubuntu 24.04. In RQ2 and RQ3, all fuzzers run in single-threaded mode 
whereas in RQ1, \toolname{} runs massively in parallel.    

\paragraph{\textbf{Metrics}}
In this evaluation, we use the following metrics:    
\begin{itemize}
    \item \emph{Validity}: the ratio of unique valid mutants over all unique mutants. 
     A mutant is considered valid if it leads to successful termination of the SUT (i.e., zero exit code). 
     Mutants leading to timeouts are counted as invalid.
     \item \emph{Yield}: 
     the ratio of typed templates whose instantiation results in a valid program 
    leading to successful termination of the SUT (i.e., zero exit code).
    \item \emph{Line coverage}: the line coverage achieved in the source code 
    of the given SUT by valid mutants.
\end{itemize}
All metrics are conditioned on valid mutants to measure the ability to deeply    
exercise program behavior. Program behavior on invalid inputs is beyond the scope    
of this paper. When reporting validity, we eliminate duplicates. 
We define yield as a metric to evaluate the effectiveness of \toolname{}'s 
preprocessing step. As a typed template whose instantiation is invalid is  
practically useless in the fuzzing loop, this metric is critical to quantify       
to what extent \toolname{} can leverage seeds at all. To record line coverage, 
we use gcov for gcc, g++, and cvc5, llvm coverage for rustc, JaCoCo 
for javac, the \texttt{coverage} pip package for Qiskit and Go's \texttt{coverpkg} 
build flag. 

\paragraph{\textbf{Time limits and seed order}}
For each controlled experimental run (RQ2, RQ3) there is 6 hours wall-clock time budget.
Within a run, \toolname{} processes seeds in random order and fuzzes each seed for 
250 iterations until the time budget is exhausted. To account for variance, 
we repeat runs of all fuzzers four times. For consistency, we ensure that all fuzzers 
using seed files have access to the same seed set and process it in the 
same order. 

\subsection*{RQ1: \rqOne}
\newlength{\buglistingindent}
\setlength{\buglistingindent}{1.5em}
\begin{figure*}[tp]
\centering
\begin{subfigure}[t]{0.47\linewidth}
\begin{minted}[xleftmargin=\buglistingindent]{c}
void a() {
  double b = b, c;
  if (c)
    c = -c;
  double d = b == c ? b : c;
  if (c != d)
    a();
}
\end{minted}
  \caption{\buglink{GCC \#125039}{GCC \#XXXX}{https://gcc.gnu.org/bugzilla/show_bug.cgi?id=125039}}
  \label{fig:bug-gallery-gcc-125039}
\end{subfigure}
\hfill
\begin{subfigure}[t]{0.47\linewidth}
\begin{minted}[xleftmargin=\buglistingindent]{common-lisp}
(declare-fun a () Int)
(declare-fun b () Int)
(assert (= (- 12) (* b (/ a (- (+ (* (-
  (cos 0.0)) 1) (cos 0.0) 0.25936 0.32))))))
(check-sat)
\end{minted}
  \caption{\buglink{Z3Prover/z3 \#10653}{Z3Prover/z3 \#YYYY}{https://github.com/Z3Prover/z3/issues/10653}}
  \label{fig:bug-gallery-z3-10653}
\end{subfigure}

\vspace{0.5em}
\begin{subfigure}[t]{0.47\linewidth}
\begin{minted}[xleftmargin=\buglistingindent]{erlang}
-module(main).
-export([main/0]).

main() -> f(1).

f(N) -> f((1 rem N) - 2).
\end{minted}
  \caption{\buglink{erlang/otp \#11400}{erlang/otp \#XXXX}{https://github.com/erlang/otp/issues/11400}}
  \label{fig:bug-gallery-erlang-11400}
\end{subfigure}
\hfill
\begin{subfigure}[t]{0.47\linewidth}
\begin{minted}[xleftmargin=\buglistingindent]{python}
import numpy as np
from qiskit.synthesis.linear_phase import \
    synth_cz_depth_line_mr
mat = np.zeros((2, 1))
qc = synth_cz_depth_line_mr(mat)
\end{minted}
  \caption{\buglink{Qiskit/qiskit \#16152}{Qiskit/qiskit \#YYYY}{https://github.com/Qiskit/qiskit/issues/16152}}
  \label{fig:bug-gallery-qiskit-16152}
\end{subfigure}

\vspace{0.5em}
\begin{subfigure}[t]{0.47\linewidth}
\begin{minted}[xleftmargin=\buglistingindent]{javascript}
const a = [0];
a.push(a);
Math.floor(a);
\end{minted}
  \caption{\buglink{jerryscript/jerryscript \#5290}{jerryscript/jerryscript \#XXXX}{https://github.com/jerryscript-project/jerryscript/issues/5290}}
  \label{fig:bug-gallery-jerryscript-5290}
\end{subfigure}
\hfill
\begin{subfigure}[t]{0.47\linewidth}
\begin{minted}[xleftmargin=\buglistingindent]{csharp}
public class Foo {
    public static bool foo()
    { return default != true; }
}
\end{minted}
  \caption{\buglink{dotnet/roslyn \#84642}{dotnet/roslyn \#YYYY}{https://github.com/dotnet/roslyn/issues/84642}}
  \label{fig:bug-gallery-roslyn-84642}
\end{subfigure}

\vspace{0.5em}
\begin{subfigure}[t]{0.47\linewidth}
\begin{minted}[xleftmargin=\buglistingindent]{go}
package main

func main() {}

var foo = (func() int {
label:
	goto label
	x := [1024 * 64]*[2]*int{}
	if x != x {_ = x}

	return 1
}())
\end{minted}
  \caption{\buglink{golang/go \#80097}{golang/go \#XXXX}{https://github.com/golang/go/issues/80097}}
  \label{fig:bug-gallery-go-80097}
\end{subfigure}
\hfill
\begin{subfigure}[t]{0.47\linewidth}
\begin{minted}[xleftmargin=\buglistingindent]{python}
import torch
from torch.autograd import Variable
w = Variable(torch.Tensor([1.0]), requires_grad=True)
def forward(x):
    return x * w
def loss(x, y):
    y_pred = forward(x)
    return (True) ** y_pred
l = loss(1.0, 1.0)
l.backward()
\end{minted}
  \caption{\buglink{pytorch/pytorch \#182283}{pytorch/pytorch \#YYYY}{https://github.com/pytorch/pytorch/issues/182283}}
  \label{fig:bug-gallery-pytorch-182283}
\end{subfigure}

\vspace{0.5em}
\begin{subfigure}[t]{0.47\linewidth}
\begin{minted}[xleftmargin=\buglistingindent]{rust}
use std::hint::black_box;
fn a(d: u8) {
    loop { 1 % d; }
}
pub fn e(d: u8) {
    if d == 0 || black_box(false) {
        a(d);
    }
}
\end{minted}
  \caption{\buglink{rust-lang/rust \#155836}{rust-lang/rust \#XXXX}{https://github.com/rust-lang/rust/issues/155836}}
  \label{fig:bug-gallery-rust-155836}
\end{subfigure}
\hfill
\begin{subfigure}[t]{0.47\linewidth}
\begin{minted}[xleftmargin=\buglistingindent]{systemverilog}
module foo(input x);
endmodule

module top
foo foo1(+++0);
endmodule
\end{minted}
  \caption{\buglink{MikePopoloski/slang \#1888}{MikePopoloski/slang \#YYYY}{https://github.com/MikePopoloski/slang/issues/1888}}
  \label{fig:bug-gallery-sv-1888}
\end{subfigure}

\vspace{0.5em}
\begin{subfigure}[t]{0.47\linewidth}
\begin{minted}[xleftmargin=\buglistingindent]{kotlin}
object Foo {
    val result = ++result
}
\end{minted}
  \caption{\buglink{Kotlin YouTrack \#88576}{Kotlin YouTrack \#XXXX}{https://youtrack.jetbrains.com/issue/KT-88576}}
  \label{fig:bug-gallery-kotlin-88576}
\end{subfigure}
\hfill
\begin{subfigure}[t]{0.47\linewidth}
\begin{minted}[xleftmargin=\buglistingindent]{php}
<?php
mb_strpos("AA\xf0\x90", "xyz", 3);
\end{minted}
  \caption{\buglink{php/php-src \#23106}{php/php-src \#YYYY}{https://github.com/php/php-src/issues/23106}}
  \label{fig:bug-gallery-php-23106}
\end{subfigure}

\vspace{0.5em}
\begin{subfigure}[t]{0.47\linewidth}
\begin{minted}[xleftmargin=\buglistingindent]{scala}
object foo {
    class ::[H, T]
    def f[A]: Unit = ()
    f[Int :: () + Int]
}
\end{minted}
  \caption{\buglink{Scala \#13190}{scala/bug \#XXXX}{https://github.com/scala/bug/issues/13190}}
  \label{fig:bug-gallery-scala-13190}
\end{subfigure}
\hfill
\begin{subfigure}[t]{0.47\linewidth}
\begin{minted}[xleftmargin=\buglistingindent]{sql}
CREATE TABLE t(b INT,c INT,d INT,e INT);
INSERT INTO t VALUES(0,0,-10,0),(0,-1,10,0);
SELECT CASE WHEN c > (SELECT avg(c) FROM t) THEN c ELSE b END FROM t WHERE c BETWEEN b AND coalesce(e,NULL) AND c > d;
\end{minted}
  \caption{\buglink{duckdb/duckdb \#26017}{duckdb/duckdb \#YYYY}{https://github.com/duckdb/duckdb/issues/26017}}
  \label{fig:bug-gallery-duckdb}
\end{subfigure}

\caption{Assorted bug samples found by \toolname{}.\anonnote{ Bug report numbers are masked to preserve anonymity.}}
\label{fig:bug-gallery}
\end{figure*}

\noindent
To assess the bug-finding capabilities of \toolname{}, we continuously ran a 
bug-hunting campaign \bughuntingMonths{} with \toolname{}. During this campaign, we have 
reported a total of \reportedBugs{} bugs, out of which \unknownBugs{} were previously undiscovered.\footnote{
Developers classified \duplicateBugs{} bug reports as duplicates.} 
Of the previously undiscovered bugs, \confirmedBugs{} are confirmed and \fixedBugs{} already fixed by 
the developers. Table~\ref{tab:bugs} shows a detailed breakdown of \toolname{}'s findings. 
We observe many bugs in SMT (Z3 and cvc5), Go (gccgo, gc), C (gcc, clang), C++ (g++, clang++) 
and SystemVerilog (slang). However, \toolname{} found 
bugs across the board: in the python libraries PyTorch and Qiskit, the official Rust compiler (rustc), a JavaScript engine, PHP, and the SQL engine DuckDB. This shows that \toolname{}
is a highly generic way of fuzzing, able to find bugs in the large majority of input languages and targets. This is particularly remarkable as the chosen languages are highly diverse with little in common.

\begin{table}[t]
\centering
\scriptsize
\caption{Bugs found with \toolname{} in a 6-month bug hunting campaign.}
\label{tab:bugs}
\begin{tabular}{llrrrr}
\toprule
Language & Target & Reported & New & Confirmed & Fixed \\
\midrule
SMT-LIB2 & cvc5 & 20 & 20 & 19 & 16 \\
 & Z3 & 20 & 19 & 18 & 10 \\
Go & gccgo & 13 & 13 & 0 & 0 \\
 & Go & 8 & 8 & 8 & 7 \\
C & GCC & 10 & 9 & 9 & 5 \\
 & Clang & 5 & 4 & 4 & 3 \\
C++ & G++ & 7 & 7 & 7 & 2 \\
 & Clang++ & 5 & 3 & 3 & 0 \\
Erlang & Erlang & 7 & 7 & 7 & 7 \\
SystemVerilog & slang & 7 & 7 & 7 & 7 \\
Python & Qiskit & 6 & 6 & 5 & 3 \\
 & PyTorch & 3 & 3 & 3 & 2 \\
 & CPython & 0 & 0 & 0 & 0 \\
PHP & PHP & 5 & 5 & 5 & 3 \\
C\# & Roslyn & 3 & 3 & 3 & 3 \\
JavaScript & JerryScript & 3 & 3 & 0 & 0 \\
Kotlin & kotlinc & 3 & 3 & 3 & 2 \\
Rust & rustc & 3 & 3 & 3 & 2 \\
Java & OpenJDK & 2 & 2 & 0 & 0 \\
Perl & Perl & 2 & 0 & 0 & 0 \\
Scala & scalac & 2 & 2 & 2 & 1 \\
SQL & DuckDB & 1 & 1 & 1 & 1 \\
 & SQLite & 0 & 0 & 0 & 0 \\
Lua & Lua & 0 & 0 & 0 & 0 \\
Solidity & solc & 0 & 0 & 0 & 0 \\
TypeScript & scriptc & 0 & 0 & 0 & 0 \\
 & tsc & 0 & 0 & 0 & 0 \\
\midrule
\textbf{Total} & & \textbf{135} & \textbf{128} & \textbf{107} & \textbf{74} \\
\bottomrule
\end{tabular}
\end{table}

\subsubsection*{\textbf{Assorted bug samples}}
The following section illustrates the diversity of bugs that \toolname{} found in     
the various input formats. For brevity, we only present bug triggers of a reasonable         
size after reduction.
\begin{description}[itemsep=2pt]
\item[Fig.~\ref{fig:bug-gallery-gcc-125039}] shows a bug in GCC: 
a cyclically initialized variable ``\texttt{b}'' and an uninitialized variable 
``\texttt{c}'' cause an internal compiler error in the range analysis. \toolname{} 
triggered this bug by substituting ``\texttt{b}'' into its own initializer.
In the original seed file, ``\texttt{b}'' was initialized with a constant.
In addition to being fixed, this bug trigger was added to the official GCC 
test suite.

\item[Fig.~\ref{fig:bug-gallery-z3-10653}] shows a soundness bug in the z3 SMT solver.
The SMT-LIB2 formula mixes integer arithmetic with the real arithmetic of the 
trigonometric operator \texttt{cos}, introduced by a mutation.
It simplifies to: there existing integers $a$ and $b$ such that $12 = b \cdot (a / 0.57936)$.
Z3 erroneously claimed the given formula was satisfiable and produced an invalid model.
The issue was quickly fixed by Z3 maintainers.

\item[Fig.~\ref{fig:bug-gallery-erlang-11400}] shows a bug in the Erlang compiler.
The bounds calculations of the remainder (\texttt{rem}) operator were incorrect, 
leading to a failed internal consistency check. 
The \texttt{rem} operator was introduced into the program via mutation. 
This issue was fixed by the maintainers.

\item[Fig.~\ref{fig:bug-gallery-qiskit-16152}] shows a crash bug in Qiskit.
The function ``\texttt{synth\_cz\_depth\_line\_mr\_inner}'' did not perform input 
validation on the shape of its input arguments, causing the Rust backend to access 
out-of-bounds elements. This issue has been fixed.

\item[Fig.~\ref{fig:bug-gallery-jerryscript-5290}] shows a crash bug in JerryScript.
Self-referential JavaScript array arguments cause \emph{Segmentation Fault} in 
several built-in \texttt{Math} functions.

\item[Fig.~\ref{fig:bug-gallery-roslyn-84642}] shows a bug in the C\# compiler.
A malformed operand \texttt{default} crashed the nullability analysis, leading 
to an assertion error. This has been fixed.

\item[Fig.~\ref{fig:bug-gallery-go-80097}] shows an internal compiler error in 
the Go compiler gc. This bug uncovered a previously unhandled state, where a variable 
``\texttt{x}'' is heap-escaped but unreachable. The heap-escape was triggered by 
a mutation in the array length. The developers identified the issue as a release 
blocker. They have since fixed the issue and also backported the fix to legacy 
versions of gc.

\item[Fig.~\ref{fig:bug-gallery-pytorch-182283}] shows an internal assertion failure 
in PyTorch's autograd. The issue was that the exponential function did not handle 
type promotion for boolean exponent bases. This was marked as a high-priority issue 
by the PyTorch maintainers and subsequently fixed.

\item[Fig.~\ref{fig:bug-gallery-rust-155836}] shows an internal compiler error in 
the Rust compiler. The trigger uncovered a high-priority bug in the dominance 
analysis, causing maintainers to add a new ``\texttt{strictly\_dominates}'' 
function in its fix. Additionally, a modified version of our reproducer 
was added to the rustc test suite.

\item[Fig.~\ref{fig:bug-gallery-sv-1888}] shows a crash bug in the Slang Verilog 
processor. The invalid usage of the prefix operator made the parser insert a missing 
identifier token which crashed the expression builder during error recovery.
This bug was fixed and our reproducer was added to the official Slang test suite.

\item[Fig.~\ref{fig:bug-gallery-kotlin-88576}] shows a crash bug in the Kotlin compiler.
The cyclical reference checker crashed on postfix operators introduced by a mutation.
This has been fixed and a modified version of the reproducer has been added to the official test suite.

\item[Fig.~\ref{fig:bug-gallery-php-23106}] shows a crash in the PHP interpreter.
The offset calculation of the Unicode string handling returned wrong offsets, leading 
to out-of-bounds reads. The specific sequence "\texttt{\textbackslash{}xf0\textbackslash{}x90}" 
was substituted into the program by a mutation on string terms. 
Maintainers greatly appreciated this bug: 
\begin{quote}
\emph{Thanks for the good catch! I just need to study the code a bit and confirm if the 
right thing to do here is to clamp or to return NULL.}
\end{quote}
The bug was subsequently fixed and a trigger was added to PHP's regression test suite.   

\item[Fig.~\ref{fig:bug-gallery-scala-13190}] shows a parser bug in the Scala compiler.
Using a malformed infix operator in a typing context led to a crash in the Scala parser.
This bug has been fixed.

\item[Fig.~\ref{fig:bug-gallery-duckdb}] shows a use-after-free vulnerability 
in the duckdb query optimizer. If the query contains a subquery used to resolve 
a \texttt{CASE} statement and the \texttt{WHERE} clause and the table is scanned with multiple 
filters, it leaves column references dangling. 
This reference is written to in a later column pruning method, leading to a write into freed memory.
We have reported the trigger to the duckdb maintainers.
After our report, a crash bug with the same underlying cause was reported, though in relation to another version branch. 
The issue is now fixed on all major version branches.

\end{description}

\rqanswer{\toolname{} is effective at discovering bugs across a diverse 
range of input languages and targets. Developers appreciate the bug reports, many of 
which led to important fixes.} 

\subsection*{RQ2: \rqTwo}
Having observed \toolname{}'s effectiveness, a natural follow-up 
question is to ask why it is effective. As a first approach at this question, we       
aim to understand the share of seed programs \toolname{} can turn into usable templates, 
i.e., the yield.
A template is usable if its instantiation is a valid program.    
We observe that \toolname{} is able to achieve an average yield 97.5 \% of 
all seeds (Table~\ref{tab:rq2-yield}, first column) ranging from 89.1 \% in 
Go to 99.7 \% in Kotlin and Perl. To understand the underlying root cause, we 
examined the effect of using the repair procedure by disabling it and keeping 
everything else unchanged (Table~\ref{tab:rq2-yield}, column \emph{No repair}).  Without 
the repair procedure, there is a sharp drop in yield (Table~\ref{tab:rq2-yield}): 
on average by 72.6\%, most in Qiskit (93.9\%) and least in TypeScript (27.3\%).
Strikingly, there is a significant drop in almost every language. As an interim conclusion, 
we can state that \toolname{} can leverage close to all seeds and that this is
disproportionately due to the repair procedure.

To further understand the effectiveness of \toolname{}, we measure 
the validity rates across all languages we fuzzed. The idea is to quantify  
the ability of \toolname{}'s mutants to reach deep program behavior.
Over a fuzzing run of six hours, \toolname{} achieves a mean validity rate of 91.5\%, 
ranging from 82.2\% in C\# to 98.1\% in SQL (see Table~\ref{tab:rq2-validity}, column \toolname{}).
Given high validity rates, we want to understand their root causes. \toolname{} 
has two ways to mitigate invalidity: non-substitutable terms addressed by calibration 
and unreliable operators addressed by Thompson sampling. To understand this more deeply, we 
disable each component in turn, leaving everything else unchanged.
Without calibration, the validity rate drops by 25.4\% on average 
(Table~\ref{tab:rq2-validity}, column \emph{-Cal}), most in Perl 
(47.8\%) and least in SQL (4.8\%). Without Thompson sampling, operators 
are sampled uniformly, leading to a drop of 40.6\% on average (column \emph{-Thom}).
Disabling both lowers the validity rate further, by 53.2\% on average (column 
\emph{-Cal-Thom}), and more than disabling either one alone in every language, 
showing that the two components are complementary. We conclude that both calibration 
and Thompson sampling are crucial for \toolname{}'s effectiveness.
Beyond the controlled experiment, we also looked at the validity rate throughout the        
fuzzing campaign of RQ1. Our results reveal an average validity rate of 
\validityRate{} across all fuzzing targets.\footnote{We measure a seven-day 
window as the mean of the validity rate of all \numTargets{} SUTs.}
 
\begin{table*}[t]
    \begin{minipage}[b]{0.49\textwidth}
        \caption{Template generation yield (\%), i.e., the share of generated templates whose instantiation is accepted by the system under test.}
        \label{tab:rq2-yield}
    \end{minipage}
    \hfill
    \begin{minipage}[b]{0.49\textwidth}
        \caption{Validity rate (\%), i.e., the share of mutants accepted by the system under test. Change when disabling calibration (-Cal), Thompson sampling (-Thom), or both (-Cal-Thom).}
        \label{tab:rq2-validity}
    \end{minipage}

    \medskip

    \begin{minipage}[t]{0.49\textwidth}
        \centering
        \scriptsize
        \renewcommand{\arraystretch}{0.9}
        \setlength{\tabcolsep}{4pt}
        \begin{tabular}{l|r!{\vrule width 1pt}rr}
            \toprule
            & \textbf{\toolname{}} & \multicolumn{2}{c}{\textbf{No repair}} \\
            \cmidrule(lr){3-4}
            Language & Yield (\%) & Yield (\%) & $\Delta$ (\%) \\
            \midrule
                C              &   96.9 &   21.4 &  -75.6 \\
                C++            &   98.1 &   21.9 &  -76.2 \\
                C\#            &   99.5 &   44.8 &  -54.7 \\
                Erlang         &   98.9 &   30.3 &  -68.6 \\
                Go             &   89.1 &    3.7 &  -85.5 \\
                Java           &   98.7 &   10.7 &  -88.0 \\
                JavaScript     &   96.9 &   11.6 &  -85.3 \\
                Kotlin         &   99.7 &   26.3 &  -73.4 \\
                Lua            &   99.4 &   12.3 &  -87.1 \\
                Perl           &   99.7 &   10.4 &  -89.3 \\
                PHP            &   98.6 &   37.6 &  -61.0 \\
                Python         &   98.4 &   16.7 &  -81.8 \\
                PyTorch        &   99.6 &   64.8 &  -34.8 \\
                Qiskit         &   94.5 &    0.6 &  -93.9 \\
                Rust           &   95.3 &    6.0 &  -89.3 \\
                Scala          &   98.9 &    9.7 &  -89.2 \\
                SMT-LIB2       &   99.4 &   28.3 &  -71.1 \\
                Solidity       &   98.5 &   34.0 &  -64.5 \\
                SQL            &   99.3 &   43.3 &  -56.0 \\
                SystemVerilog  &   97.4 &   24.3 &  -73.1 \\
                TypeScript     &   91.6 &   64.3 &  -27.3 \\
            \midrule
                \textbf{Average} &   97.5 &   24.9 &  -72.6 \\
            \bottomrule
        \end{tabular}
    \end{minipage}
    \hfill
    \begin{minipage}[t]{0.49\textwidth}
        \centering
        \scriptsize
        \renewcommand{\arraystretch}{0.9}
        \setlength{\tabcolsep}{4pt}
        \begin{tabular}{l|r!{\vrule width 1pt}rrr}
            \toprule
            & \textbf{\toolname{}} & \textbf{-Cal} & \textbf{-Thom} & \textbf{-Cal-Thom} \\
            Language & Validity (\%) & $\Delta$ (\%) & $\Delta$ (\%) & $\Delta$ (\%) \\
            \midrule
                C              &   93.7 &  -29.2 &  -43.4 &  -54.5 \\
                C++            &   91.1 &  -29.3 &  -47.1 &  -60.8 \\
                C\#            &   82.2 &  -19.1 &  -47.6 &  -52.8 \\
                Erlang         &   92.3 &  -17.2 &  -38.9 &  -48.9 \\
                Go             &   94.0 &  -26.4 &  -57.4 &  -66.9 \\
                Java           &   89.7 &  -29.9 &  -52.7 &  -59.1 \\
                JavaScript     &   93.1 &  -27.3 &  -18.2 &  -42.8 \\
                Kotlin         &   86.6 &  -29.7 &  -43.2 &  -53.2 \\
                Lua            &   91.5 &  -38.9 &  -26.4 &  -51.4 \\
                Perl           &   85.6 &  -47.8 &  -36.5 &  -59.1 \\
                PHP            &   95.3 &  -22.1 &  -29.8 &  -45.4 \\
                Python         &   82.3 &  -27.6 &  -24.2 &  -38.5 \\
                PyTorch        &   90.4 &  -26.3 &  -27.5 &  -49.8 \\
                Qiskit         &   93.0 &  -11.6 &  -15.7 &  -28.5 \\
                Rust           &   93.0 &  -30.4 &  -42.5 &  -62.5 \\
                Scala          &   91.2 &  -27.1 &  -60.7 &  -65.4 \\
                SMT-LIB2       &   89.9 &  -26.7 &  -45.8 &  -53.3 \\
                Solidity       &   96.6 &  -20.5 &  -71.2 &  -76.5 \\
                SQL            &   98.1 &   -4.8 &  -44.7 &  -45.7 \\
                SystemVerilog  &   95.3 &  -24.5 &  -35.2 &  -49.1 \\
                TypeScript     &   96.4 &  -16.2 &  -44.9 &  -52.8 \\
            \midrule
                \textbf{Average} &   91.5 &  -25.4 &  -40.6 &  -53.2 \\
            \bottomrule
        \end{tabular}
    \end{minipage}
\end{table*}

\rqanswer{The effectiveness of \toolname{} can be explained by high template yield 
and validity rates. High yield is achieved by the repair procedure. 
Additionally, calibration and Thompson sampling significantly improve validity rates.}

\subsection*{RQ3: \rqThree}
\begin{figure}[t]
\input{includes/rq2-graphs.tex}
\bigskip
    \begin{minipage}[b]{0.52\textwidth}
        \captionof{table}{Total programs generated, valid programs, validity rate, and line coverage per language.
        All programs were deduplicated.}
        \label{tab:rq2-full}
    \end{minipage}
    \hfill
    \begin{minipage}[b]{0.46\textwidth}
        \captionof{table}{Validity rate (\%) of domain-specific fuzzers and difference $\Delta$ (percentage points) of \toolname{} to them.}
        \label{tab:rq2-traditional}
    \end{minipage}

    \medskip

    \begin{minipage}[t]{0.52\textwidth}
        \centering
        \scriptsize
        \setlength{\tabcolsep}{3pt}
        \begin{tabular}[t]{llrrrr}
            \toprule
            Target & Fuzzer & Programs & Valid & Validity & Lines Cov. \\
            \midrule
            \multirow{3}{*}{GCC} & Fuzz4All & 13,854 & 4,787 & 34.6\% & 325,628 \\
             & Kitten & 739,851 & 123,488 & 16.7\% & 397,120 \\
             & \toolname{} & 307,222 & \textbf{297,460} & \textbf{96.8\%} & \textbf{420,333} \\
            \addlinespace[5pt]
            \multirow{3}{*}{G++} & Fuzz4All & 8,002 & 2,131 & 27.5\% & 376,179 \\
             & Kitten & 37,281 & 1,295 & 3.5\% & 328,688 \\
             & \toolname{} & 27,816 & \textbf{25,977} & \textbf{93.4\%} & \textbf{396,036} \\
            \addlinespace[5pt]
            \multirow{3}{*}{go} & Fuzz4All & 13,583 & 4,502 & 33.7\% & 113,729 \\
             & Kitten & 712,377 & 16,566 & 2.3\% & 114,022 \\
             & \toolname{} & 151,960 & \textbf{145,942} & \textbf{96.0\%} & \textbf{115,622} \\
            \addlinespace[5pt]
            \multirow{3}{*}{javac} & Fuzz4All & 11,655 & 5,419 & 46.7\% & \textbf{14,098} \\
             & Kitten & 70,728 & 10,524 & 14.9\% & 13,078 \\
             & \toolname{} & 43,573 & \textbf{39,800} & \textbf{91.3\%} & 13,716 \\
            \addlinespace[5pt]
            \multirow{2}{*}{Qiskit} & Fuzz4All & 10,926 & 4,308 & 39.1\% & 14,552 \\
             & \toolname{} & 65,480 & \textbf{63,874} & \textbf{97.5\%} & \textbf{15,774} \\
            \addlinespace[5pt]
            \multirow{3}{*}{rustc} & Fuzz4All & 10,013 & 2,688 & 27.0\% & 152,869 \\
             & Kitten & 267,732 & 24,768 & 9.3\% & 176,033 \\
             & \toolname{} & 106,522 & \textbf{100,939} & \textbf{94.8\%} & \textbf{176,116} \\
            \addlinespace[5pt]
            \multirow{3}{*}{cvc5} & Fuzz4All & 12,549 & 6,156 & 49.1\% & 67,077 \\
             & Kitten & 275,124 & \textbf{59,885} & 21.8\% & 87,569 \\
             & \toolname{} & 58,102 & 53,224 & \textbf{91.5\%} & \textbf{91,399} \\
            \bottomrule
        \end{tabular}
    \end{minipage}
    \hfill
    \begin{minipage}[t]{0.46\textwidth}
        \centering
        \scriptsize
        \setlength{\tabcolsep}{3pt}
        \renewcommand{\arraystretch}{1.3}
        \begin{tabular}[t]{llrr}
            \toprule
            Target & Domain-specific fuzzer & Validity & $\Delta$ \toolname{} \\
            \midrule
            GCC & Csmith & \textbf{99.96} & $-6.22$ \\
            G++ & YARPGen & \textbf{97.93} & $-6.82$ \\
            csc & Fuzzlyn & \textbf{98.98} & $-16.82$ \\
            erl & erlfuzz & 91.25 & $\mathbf{+1.09}$ \\
            go & GoSmith & \textbf{99.85} & $-5.88$ \\
            javac & Hephaestus & \textbf{97.10} & $-7.36$ \\
            kotlinc & Kotlin Compiler Fuzzer & \textbf{92.72} & $-6.08$ \\
            PyTorch & DeepREL & \textbf{99.69} & $-9.29$ \\
            Qiskit & MorphQ & 72.00 & $\mathbf{+21.02}$ \\
            rustc & Rustlantis & \textbf{100.00} & $-7.00$ \\
            cvc5 & TypeFuzz & 67.29 & $\mathbf{+22.63}$ \\
            solc & Erwin & \textbf{99.98} & $-3.38$ \\
            SQLite & SQLsmith & 49.20 & $\mathbf{+48.85}$ \\
            slang & Verismith & \textbf{100.00} & $-4.74$ \\
            \bottomrule
        \end{tabular}
    \end{minipage}

\end{figure}

As a follow-up question, we ask how \toolname{} compares to other fuzzers.
To this end, we first compare it against Fuzz4All~\cite{xia-etal-icse24}, 
the current state-of-the-art LLM-based fuzzer, and Kitten~\cite{xie-etal-issta25}, 
a grammar-based fuzzer proposed as a baseline for LLM-based fuzzers. 
Like \toolname{}, both are black-box fuzzers that target a wide range of formats.
We compare them in terms of validity rate and line coverage. We run all three fuzzers on the targets used by Fuzz4All (GCC, G++, Go, javac, 
Qiskit and cvc5), and on rustc, for which we use the Fuzz4All configuration from Kitten's evaluation~\cite{xie-etal-issta25}.
Kitten requires an ANTLR grammar, which is not available for Qiskit, so we omit Kitten there.
To fit Fuzz4All into the GPU memory of our machine, we reduced its batch size to 5 programs per LLM query.

We observe that \toolname{} maintains a validity rate above 90\% for all targets. 
By contrast, the validity rate of Fuzz4All stays below 50\% across targets, and that of Kitten below 25\%, dropping to 3.5\% in G++ and 2.3\% in Go (Table~\ref{tab:rq2-full}).
In terms of coverage, \toolname{} improves upon the baselines in GCC, G++, 
Qiskit and cvc5, is on par in Go and rustc, and shows slightly lower coverage 
than Fuzz4All in javac. Looking at coverage over time (Fig.~\ref{fig:rq2-graphs}), 
\toolname{} demonstrates similar growth as Fuzz4All, whereas Kitten saturates 
more quickly. Notably, \toolname{} leads on 6 out of 7 targets.

To put the validity rate of \toolname{} into perspective, we also compare to 
domain-specific fuzzers. Where one exists, we select the state-of-the-art domain-specific fuzzer: 
Csmith \cite{yang-etal-pldi11} for C, YARPGen \cite{livinskii-etal-oopsla20} for C++, Fuzzlyn \cite{jakobbotsch-github26} for C\#, erlfuzz \cite{WhatsApp-github26} for Erlang, GoSmith \cite{dvyukov-github26} for Go, Hephaestus \cite{chaliasos-pldi22} for Java, Kotlin Compiler Fuzzer \cite{Georgescu-fse24} for Kotlin, DeepREL \cite{deng-fse22} for PyTorch, MorphQ \cite{Paltenghi-icse23} for Qiskit, Rustlantis \cite{wang-oopsla24} for Rust, SQLsmith \cite{anse1-github26} for SQL, Verismith \cite{Herklotz-fpga20} for SystemVerilog, Erwin \cite{ma-icse26} for Solidity and the mutation-based fuzzer TypeFuzz \cite{park-etal-oopsla21} for SMT-LIB2. 
Most achieve near-perfect validity, several by construction (Table~\ref{tab:rq2-traditional}).
Csmith, a domain-specific fuzzer at more than 40k LoC, achieves 99.96\%.
Despite deriving all language knowledge from templates, \toolname{} stays within a 
few percentage points of most of the domain-specific fuzzers, with the largest gap 
in C\# ($-16.82$ percentage points vs.\ Fuzzlyn). Notably, \toolname{} has a higher validity 
rate than erlfuzz, MorphQ, TypeFuzz and SQLsmith. TypeFuzz uses the same corpus as 
\toolname{}, but supports only a subset of all SMT-LIB2 files, 
producing invalid mutants for the others.
SQLsmith primarily targets PostgreSQL and emits constructs that our target SQLite rejects.
Both cases highlight that domain-specific fuzzers lose validity if targets diverge from 
the format they were designed for.

\rqanswer{\toolname{} achieves a categorically higher 
validity rate than Fuzz4All and Kitten, and performs noticeably higher or is on par 
in terms of coverage. Its validity rate comes close to that of dedicated domain-specific fuzzers and 
exceeds several of them.}

\paragraph{\textbf{Threats to validity}}
The main internal threat is the correctness of \toolname{}'s
implementation. To mitigate it, \toolname{} ships with a test suite. For
comparison, we run each baseline from its publicly available implementation,
and we give all seed-based fuzzers the same seed set in the same order.
The main external threat is the choice of evaluation targets.
To support our generality claim, we evaluate \toolname{} on \numTargets{}
targets across \numInputLanguages{} diverse input languages. To account for variance,
we repeat each 6-hour run of RQ2 and RQ3 four times. Unlike fuzzers that query
an LLM throughout fuzzing, \toolname{} uses an LLM only in preprocessing. Its
fuzzing loop is therefore unaffected by LLM non-determinism once the 
templates exist. LLM hallucinations can still affect the
templates. Repair can fix mistakes outside terms and drops terms that do not
occur in the seed, and calibration strips non-substitutable terms. 

\section{Related Work}
\label{sec:related-work}
We survey related work on four strands: (1) LLM-based fuzzing, (2) template-based
fuzzing, (3) type-aware and domain-specific fuzzing and (4) domain-agnostic fuzzing.

\paragraph{\textbf{LLM-based fuzzing}}
Our methodology is related to LLM-based fuzzing. We differentiate two families:
using LLMs for test generation directly (i.e., in the fuzzing loop) and 
using LLMs to generate fuzzer machinery. Of the first family, the earliest approach 
is TitanFuzz \cite{deng-etal-issta23} 
using LLMs for testing deep-learning libraries. About a year later, 
Fuzz4All \cite{xia-etal-icse24} demonstrated that LLMs are effective  
at language-agnostic test generation by finding bugs in various targets. 
More recent approaches further augmented LLM-based test generation by inspecting source 
code~\cite{yang-etal-oopsla24}, historical bug reports~\cite{deng-etal-icse24}
or by using LLM calls to repeatedly fill gaps instead of generating
the tests from scratch~\cite{gao-icse25}. Their shared shortcoming is the excessive number
of LLM calls in the fuzzing loop.
Instead of using LLMs for test generation directly, a second family of approaches 
uses LLMs to generate fuzzer machinery. Most closely related 
is LegoFuzz \cite{ni-etal-oopsla25}, an approach for testing C compilers that 
uses LLMs in a preprocessing step to produce a corpus of self-contained functions.
Another approach, MetaMut \cite{ou-etal-asplos25}, combines LLMs with expert 
knowledge to write AST-manipulation code for coverage-guided fuzzing of C 
compilers. Once4All \cite{sun-etal-asplos26} lets the LLM write generators 
to fill formula skeletons for testing SMT solvers. 
Most recently, GenToo \cite{vikram-ase26} was proposed. GenToo is a fully 
agentic approach which autonomously refines input generators for Java libraries. 
A key benefit that this approach family shares with TTF is to enable LLMs to     
escape the fuzzing loop. However, in contrast
to \approachname{}, these approaches rely on language-specific tooling, 
LLM refinement calls, human correction, or frontier model access.   

\paragraph{\textbf{Template-based fuzzing}}
\approachname{} is related to template-based fuzzing. The fuzzers JAttack 
and LeJit \cite{zang-etal-ase23, zang-etal-fse24} use templates to test the Java JIT. 
JAttack relies on developer-written templates whereas LeJit \cite{zang-etal-fse24} 
derives templates from existing programs and uses the Java toolchain for typing. 
TemuJS~\cite{wong-oopsla25} uses templates to test JavaScript engines.
Its key idea is abstracting programs into templates with JavaScript-specific placeholders. 
Similar to TTF, these approaches abstract programs with templates.
However, they require either 
human expertise or domain-specific tooling for template generation,
limiting these approaches to a few languages. By contrast, TTF uses an LLM to create
templates, enabling a wide range of input languages at little effort.
At a high level, \approachnameShorthand{} is related to program sketching \cite{solarlezama-phd08},
an approach for program synthesis where partial programs are populated 
with expressions under typing and scoping constraints.

\paragraph{\textbf{Type-aware and domain-specific fuzzing}}
\approachname{} is related to type-aware fuzzing.   
Instantiated on SMT, \toolname{}'s mutation strategy is similar to that of      
TypeFuzz \cite{park-etal-oopsla21}, which performs type-aware generative mutations with 
s-expression operators. Different from TypeFuzz, \toolname{}
extends generative type-aware mutation to arbitrary input formats and
does not require a type checker.
Loosely related to our approach is SPE \cite{zhang-acm17}, which fuzzes programs with
holes, stripping C programs to skeletons where variable occurrences are replaced 
with holes. It then performs exhaustive enumeration of all variable usage patterns 
in the program. Besides variables, \approachnameShorthand{} substitutes 
constants and nested expressions and employs sampling instead of enumeration. 

\paragraph{\textbf{Domain-agnostic fuzzing}}
\approachname{} shares the aim of domain-agnostic fuzzing predating the LLM era: 
to be broadly applicable across languages.     
Tools from the AFL family~\cite{zalewski-web16, fioraldi-etal-woot20,aschermann-etal-ndss19,groce-etal-cc22} 
use generic bit-level mutations and coverage guidance to be applicable generically. Unlike TTF,    
most mutants of AFL-based fuzzers are invalid. Another stream of domain-agnostic    
fuzzing is grammar-based fuzzing. Grammarinator \cite{hodovan-etal-atest18} is a 
black-box fuzzer which uses user-provided ANTLR grammars. Kitten \cite{xie-etal-issta25} 
also relies on ANTLR grammars, beating Fuzz4All in a code coverage comparison on 
a few targets. Kitten hence serves as a baseline for LLM-based fuzzing approaches. 
\toolname{} outperforms Kitten in code coverage on valid mutants. 
Another grammar-based fuzzing approach is language-based fuzzing~\cite{steinhofel-etal-fse22} which 
augments context-free grammars with language constraints 
to generate targeted inputs for fuzzing. While TTF and this approach both        
target semantically valid inputs, language-based fuzzing requires a grammar 
and a specification for each language including type semantics and scoping rules. 
By contrast, TTF is lightweight and solely requires
transforming test cases into templates. This enables \toolname{}'s adaptation to \numInputLanguages{} 
input languages at validity rates close to those of domain-specific fuzzers.

\section{Conclusion and Future Work}
\label{sec:conclusion}
We propose \approachname{}, a lightweight LLM-based fuzzing methodology.
The key idea is to use an LLM only in preprocessing: to transform fuzzing seeds
into typed templates, which are then used for fuzzing.
As a key benefit, \approachnameShorthand{} is flexible, needing no
domain-specific tooling such as parsers or type checkers.
Moreover, the correspondence between templates and seeds enables repairing LLM mistakes.
We engineered \toolname{}, a realization of \approachname{}.
Using \toolname{}, we conducted an extensive bug-hunting campaign on \allTargets{}.
We found \unknownBugs{} new bugs, of which \fixedBugs{} have already been fixed.
Perhaps most notably, TTF yields \validityRate{} of valid mutants across 
all input languages.
Our evaluation further showed unique benefits over, and orthogonality to, existing tools.
In the future, we aim to investigate whether \approachnameShorthand{} can be extended
to find logic and instability bugs via declarative metamorphic oracles.
Equality saturation and e-graphs~\cite{willsey-etal-popl21} offer a
promising avenue to constrain \toolname{}'s mutations to be semantics-preserving. 

\section*{Data Availability}\phantomsection\label{sec:data-availability}
The source code for \toolname{}, the seed files, prompts, experimental data and evaluation scripts are available at \url{https://doi.org/10.5281/zenodo.23079984}.
\bibliographystyle{ACM-Reference-Format}
\bibliography{bibliography}

\end{document}